\documentclass[structabstract]{aa}
\usepackage{graphicx}
\usepackage{longtable,lscape}
\usepackage{txfonts}
\begin{document}
   \title{A likely exoplanet orbiting the oscillating K-giant $\alpha$ Arietis}

   \author{Byeong-Cheol Lee\inst{1},
          David E. Mkrtichian\inst{2},
          Inwoo Han\inst{1},
          Kang-Min Kim \inst{1},
          \and
          Myeong-Gu Park\inst{3}
          }

   \institute{Korea Astronomy and Space Science Institute, 776,
		Daedeokdae-Ro, Youseong-Gu, Daejeon 305-348, Korea\\
	      \email{[bclee;iwhan;kmkim]@kasi.re.kr}
        \and
             Crimean Astrophysical Observatory,
             Nauchny, Crimea, 98409, Ukraine\\
             \email{davidm@crao.crimea.ua}
	    \and
	      Department of Astronomy and Atmospheric Sciences,
	      Kyungpook National University, Daegu 702-701, Korea\\
	      \email{mgp@knu.ac.kr}
             }

   \date{Received 10 December 2010 / Accepted 10 March 2011}


  \abstract
   {}
   {To search for extrasolar planets around intermediate-mass stars,
   we are conducting a precise RV survey around K-giants.
   }
   {We present high-accuracy RV measurements of $\alpha$ Ari made from November 2003 to February 2010.
   This star belongs to our sample of 55 K-giants studied for extrasolar planet and pulsation searches
   using the fiber-fed Bohyunsan Observatory Echelle Spectrograph (BOES) attached to the 1.8-m telescope
   at Bohyunsan Optical Astronomy Observatory (BOAO) in Korea.
   }
   {We find a planetary companion with long-period and low-amplitude radial velocity (RV) variations
   in oscillating K2 III star $\alpha$ Ari (HD 12929). We do not find the correlation between RV variations and chromospheric activity indicators (Ca II H \& K region, H$\alpha$ line).
   The bisector analysis also shows that the bisector velocity span (BVS) does not show any obvious correlation with RV variation but has periodic component that may be attributed to the rotation of the star.
   If the RV variation is indeed caused by a planetary companion,
   an orbital solution with a period of $P$ = 381 days, a semi-amplitude of $K$ = 41 m s$^{-1}$,
   and an eccentricity of $e$ = 0.25 fits the data best. Assuming a possible stellar mass of $M_{\star}$ = 1.5 $M_{\odot}$,
   we estimate the minimum mass for the planetary companion of $m_{2}$ sin $i$ = 1.8 $M_{Jup}$ with an orbital semi-major axis of 1.2 AU.
   }
   {Our finding of a likely exoplanet in $\alpha$ Ari supports searching for extrasolar planets
   around giant stars with multiperiodic oscillations.
   }


   \authorrunning{Byeong-Cheol Lee et al.}
   \titlerunning{A likely exoplanet orbiting the oscillating K-giant $\alpha$ Arietis}
   \maketitle

%

\section{Introduction}

  Ever since the first extrasolar planet around a main-sequence star was
discovered by Mayor $\&$ Queloz (1995),  about 500 extrasolar planets have been found so far.
Most of them are found around main-sequence stars using the precise RV method,
but several planets have only been found around giant stars.

Detection around intermediate-mass stars evolving toward the giant stage
is easier than around early stage or main-sequence stars thanks to the availability of
many sharp absorption lines for high-precision RV measurements.
Thus makes giant stars suitable targets for extrasolar planet detection with the RV method.
Frink et al. (2002) discovered the first planetary companion around
the K-giant star $\iota$ Dra (K2 III), and thereafter, several companions
around K-giant stars have been reported using the precise RV method
(Setiawan et al. 2003a, 2003b; Mitchell et al. 2003; Hatzes et al. 2005, 2006; Reffert et al. 2006;
Johnson et al. 2007, 2008; Do$\ddot{}$llinger et al. 2007, 2009; de Medrios et al. 2009;
and  Sato et al. 2007, 2008a, 2008b, 2010).
However, in K-giants the velocity variations caused by planetary companions
can be blended with the stellar pulsations and surface activities,
which complicates identification of planetary companions.

K-giant $\alpha$ Ari (HD 12929, HR 617) is one of our sample 55 K-giant stars
for which we have obtained precise RV measurements in the past eight
years at BOAO in Korea. In previous publication, we (Kim et al. 2006) reported
multiperiodic oscillations in $\alpha$ Ari with periods $P_{1}$ =
0.571 day (or aliases at 0.445 or 0.871 days) observed during November to
December 2003. The average amplitude variation ranges $\sim \pm$ 20 m s$^{-1}$
just overnight. This variability can only be the result of
stellar oscillations because of the short time scale. We continued to
observe $\alpha$ Ari, by getting several spectra every night to search
for long-term variation. Here we report the detection of a long-period and low-amplitude
RV variation of $\alpha$ Ari possibly due to a planetary companion.


\section{Observations and analysis}

 We acquired 678 spectra of $\alpha$ Ari from November
2003 to February 2010 using the fiber-fed high resolution (R = 90 000
at 5000 {\AA}) echelle spectrograph BOES (Kim et al. 2007) attached to
the 1.8-m telescope at BOAO in Korea. An iodine absorption cell (I$_{2}$)
was used to provide the precise RV measurements. Each estimated
signal-to-noise ratio (S/N) at I$_{2}$ wavelength region is about 250
with typical exposure time ranging between 90 and 180 seconds.

The extraction of normalized 1--D spectra was carried out using
IRAF (Tody 1986) software. Further data processing, I$_2$ analysis and
precise RV measurements were undertaken using a code called RVI2CELL (Han
et al. 2007) developed at the Korea Astronomy $\&$ Space Science Institute
(KASI).

\begin{figure}
    \centering
    \includegraphics[width=8cm]{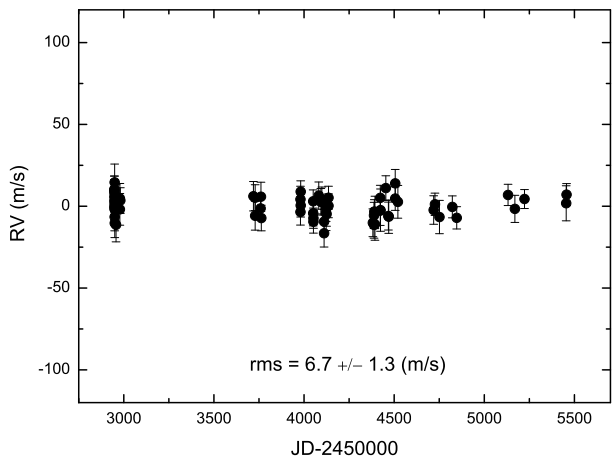}
      \caption{RV measurements of a standard star $\tau$ Ceti from 2003 to 2010.
          }
     \label{tauCeti}
   \end{figure}

To demonstrate a long-term stability of the BOES we show
relative RV measurements of the standard star $\tau$ Ceti
in Figure ~\ref{tauCeti}. It shows that RV of $\tau$ Ceti is constant with
an rms scatter of 6.7 m s$^{-1}$, over the time span of our observations.

%


\section{The properties of $\alpha$ Ari}

We refer to the main photometric parameters of $\alpha$ Ari to Allende Prieto \& Lambert
(1999) taken from the Hipparcos catalogs of stars within 100 pc from
the Sun and the main spectroscopic parameters to Hekker and Mel{\'e}ndez
(2007) taken at Lick observatory.

The mean RV of $\alpha$ Ari is  -- 14.6 km s$^{-1}$ (Evans 1967).
Berdyugina et al. (1991) derived $T_{\mathrm{eff}}$ = 4850 $\pm$ 150 K.
And, Gray et al. (2002) determined $T_{\mathrm{eff}}$ = 4481 $\pm$ 11 K
and [Fe/H] = -- 0.221 $\pm$ 0.036. Mozurkewich et al. (1991) measured an
angular diameter for $\alpha$ Ari of 6.412 $\pm$ 0.064 mas at 800 nm.
The Hipparcos parallax of 49.48 $\pm$ 0.99 mas results in a stellar
radius of 13.9 $\pm$ 0.3 $R_{\odot}$. The most recent determinations of
the  projected rotational velocity $v_{rot}$ sin $i$ = 3.44 km s$^{-1}$
by Hekker \& Mel{\'e}ndez (2007) and 3.1 $\pm$ 1 km s$^{-1}$ by Carney
et al. (2008) are close to each other. Based on the latter value that is
given with error estimation, we derived the range of upper limit for the
rotational period of

   $P_{rot} = 2 \pi$R / ($v_{rot}$ sin $i$) = 172 -- 335 days.\\

\hskip -15pt
Berdyugina et al. (1991) determined log $\it g$ = 2.6 $\pm$ 0.3,
almost in agreement with log $\it g$ = 2.57 $\pm$ 0.2 as determined by
McWilliam (1990). Surface gravity and its uncertainty log $\it g$ = 2.6
$\pm$ 0.3 yields a stellar mass in the range of 1.4 -- 5.6 $M_{\odot}$.
To estimate the mass of $\alpha$ Ari independently, we determined
atmospheric parameters directly from our spectra.
By using measured 268 equivalent widths (EW) of Fe I and Fe II
lines, we estimated $T_{\mathrm{eff}}$,
[Fe/H] = -- 0.09 $\pm$ 0.03, log $\it g$, and $v_{micro}$
of the star using the program TGVIT (Takeda et al. 2005). Our estimated
surface gravity with its formal uncertainty is log $\it g$ = 2.33 $\pm$
0.06. That yields a stellar mass of 1.5 $\pm$ 0.2 $M_{\odot}$.
Table 1 summarizes the basic stellar parameters of $\alpha$ Ari.

\begin{table}
\begin{center}
\caption[]{Stellar parameters for $\alpha$ Ari.}
\label{tab1}
\begin{tabular}{lcc}
\hline
\hline
    Parameter          & Value      &    Reference     \\

\hline
    Spectral type            & K2 III     &  Hekker \& Mel{\'e}ndez (2007) \\
    $\textit{$m_{v}$}$ [mag]  & 2.01      & Hipparcos  \\
    $\textit{$M_{v}$}$ [mag]  & 0.48 $\pm$ 0.09   & Hipparcos \\
    $\textit{B-V}$ [mag]     & 1.151     & Hipparcos \\
    Parallax [mas]           & 49.48 $\pm$ 0.99  & Hipparcos \\
    Diameter [mas]   & 6.412 $\pm$ 0.064 & Hipparcos \\
    Distance [pc]           & 20.2      & Hipparcos \\
    RV [km s$^{-1}$]        & -- 14.6 $\pm$ 0.1 & Evans (1967) \\
    $T_{\mathrm{eff}}$ [K]  & 4480           & McWilliam (1990) \\
                            & 4850 $\pm$ 150 & Berdyugina et al. (1991) \\
                            & 4481 $\pm$ 11  & Gray et al. (2002) \\
                            & 4553 $\pm$ 15  & This work \\
    $\mathrm{[Fe/H]}$       & -- 0.25  & McWilliam (1990) \\
                            & -- 0.22 & Mozurkewich et al. (1991) \\
                            & -- 0.09 $\pm$ 0.03 & This work \\
    log $\it g$             & 2.57 $\pm$ 0.2  & McWilliam (1990) \\
                            & 2.6 $\pm$ 0.3   & Berdyugina et al. (1991) \\
                            & 2.33 $\pm$ 0.06 & This work \\
    $\textit{$M_{\star}$}$ [$M_{\odot}$] & 1.5 $\pm$ 0.2 & This work \\
    $\textit{$R_{\star}$}$ [$R_{\odot}$] & 13.9 $\pm$ 0.3  & derived \\
    $v_{rot}$ sin $i$ [km s$^{-1}$] & 3.44  & Hekker \& Mel{\'e}ndez (2007) \\
                                    & 3.1  $\pm$ 1  & Carney et al. (2008) \\
    $P_{rot}$ / sin $i$ [days]      & 172 -- 335     & derived \\
    $v_{micro}$ [km s$^{-1}$]       & 1.7 & McWilliam (1990) \\
                        & 1.40 $\pm$ 0.06 & This work \\

\hline

\end{tabular}
\end{center}
\end{table}
%


\section{Orbital solution}

    The long-term variation of $\alpha$ Ari was first noted by Walker et
al. (1989) and Larson et al. (1999) who used CFHT 3.6-m and DAO
telescopes with a hydrogen-fluoride (HF) absorption cell. These data were
insufficient (45 spectra) to draw any conclusions about possible
periodic variability.  We recalculated the standard deviation of their
45 measurements of RV variation of $\sigma_{\mathrm{CFHT}}$ = 41.6 $\pm$
6.2 m s$^{-1}$ during 13 years. The standard deviation of our
678 measurements spanning eight years is  $\sigma_{\mathrm{BOAO}}$ = 27.1 $\pm$ 1.6 m
s$^{-1}$.

\begin{figure*}
   \centering
   \includegraphics[width=17cm]{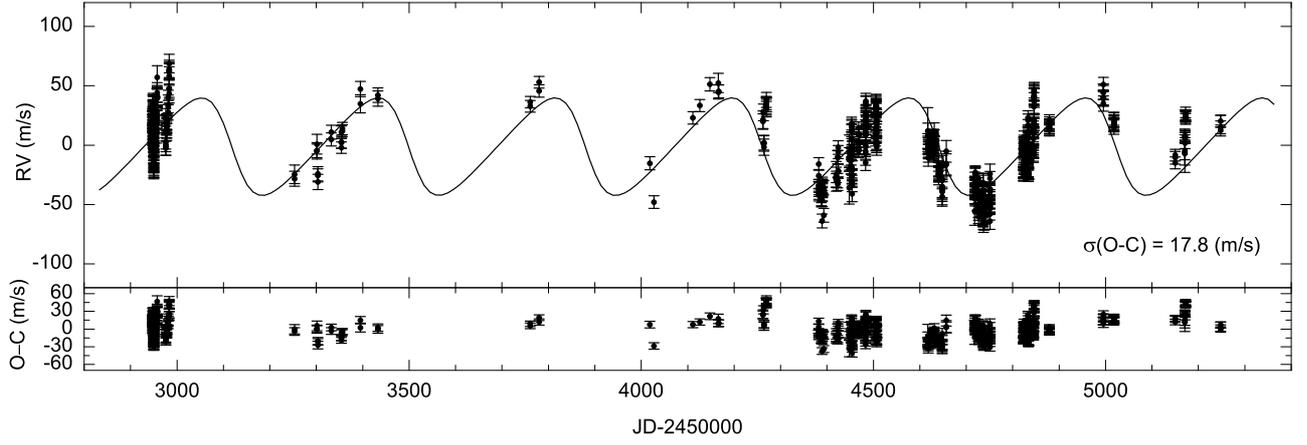}
      \caption{RV measurements (Top) and rms scatter of the residuals (Bottom)
      for $\alpha$ Ari taken at BOAO from November 2003 to February 2010.
      The solid line is the orbital solution with a period of 380.8 days
      and an eccentricity of 0.25.
          }
     \label{orbit}
   \end{figure*}

\begin{figure}
   \centering
   \includegraphics[width=8cm]{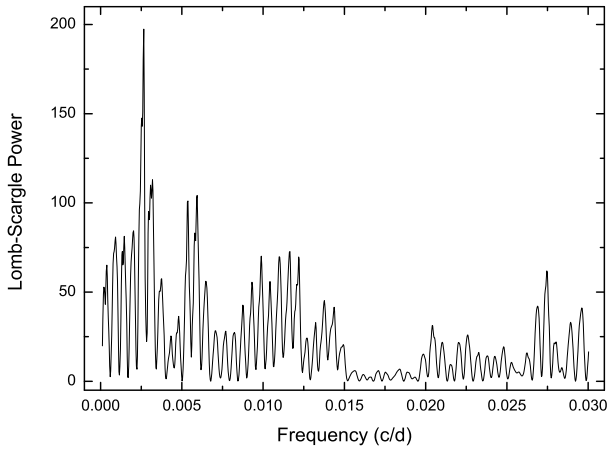}
      \caption{The Lomb-Scargle periodogram of the RV measurements for $\alpha$ Ari.
      The periogram shows a significant power at a frequency of 0.00263 c d$^{-1}$
      corresponding to a period of 380.8 days.
          }
     \label{power}
   \end{figure}

Figure~\ref{power} shows the Lomb-Scargle periodogram (Lomb 1976; Scargle 1982)
of the original RV time series for $\alpha$ Ari. It shows a significant power
at $f_{1}$=0.00263 c\,d$^{-1}$ ($P$ = 380.8 days). The individual RV
measurements are shown on the top panel in Figure~\ref{orbit}, and they
exhibit clear variations.

The statistical significance of this signal
was calculated using the bootstrap randomization technique (e.g. Hatzes \&
Mkrtichian 2004 and references herein). The measured RV values were
randomly shuffled keeping the observed times fixed.
The Lomb-Scargle periodogram was then computed 200 000 times for each ``random''
data set. The fraction of a periodogram having power higher than
the amplitude peak in the range of 0 $<$ $f$ $<$ 0.03 c d$^{-1}$
represents the false-alarm probability (FAP) that noise would create
the detected signal. A small FAP is therefore an indication of a significant peak.
We found that among 200 000 trials the highest peak
of periodograms never exceed a signal in the original data, thus the FAP
$<$ 5 $\times$ $10^{-6}$.

\begin{figure}
   \centering
   \includegraphics[width=8cm]{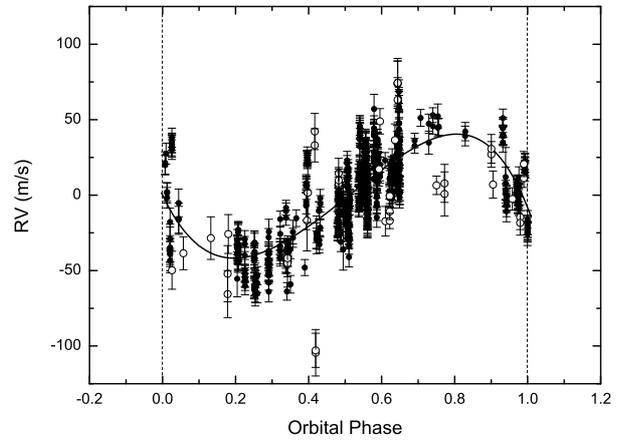}
      \caption{RVs of $\alpha$ Ari phased to the period of 380.8 days.
      Filled and open circles denote BOES and CFHT spectroscopic measurements, respectively.
      The solid line is the orbital solution that fits the data with an rms of 17.8 m s $^{-1}$.
          }
     \label{phase}
   \end{figure}

The solid line in the top panel of Figure~\ref{orbit} is the best-fit
Keplerian orbit with $P$ = 380.8 $\pm$ 0.3 days, semi-amplitude $\it K$
= 41.1 $\pm$ 0.8 m s$^{-1}$ and eccentricity of 0.25 $\pm$ 0.03. We
estimated the minimum mass of the companion to be 1.8 $\pm$ 0.2 $M_{Jup}$
at a distance of 1.2 AU from $\alpha$ Ari. The RV phase diagram for the orbit in
Figure~\ref{orbit} is shown in Figure~\ref{phase}.
CFHT data (open circles) are also shown in the phase diagram to the period of BOES data set.
Uncertainties of CFHT RV measurements are 2 $\sim$ 3 times larger than that of BOES
and also do not conform perfectly with BOES phase diagram because of low instrumental exquisiteness
and some noise pattern of the data (Larson et al. 1993, 1999).
  The rms residual from the orbital solution is 17.8 m s$^{-1}$
(Figure~\ref{orbit}). This is almost certainly due to the short-period
variations and intrinsic variability of stellar oscillations, as shown
by Kim et al. (2006). All the orbital elements are listed in Table 2.

%


\section{The nature of the RV variations}

 K-giant stars exhibit pulsation, as well as surface activity, resulting
in low-amplitude RV variability on different time scales. While
short-term period (hours $\sim$ days) RV variations have been known to be
the result of stellar pulsations (Hatzes \& Cochran 1998), long-term period
(a hundred of days) RV variations with a low-amplitude may be caused by
three kinds of phenomena: planetary companion, rotational
modulation of surface features, or non-radial oscillation modes. To
establish whether a low amplitude and long-period RV variations are
caused by surface activities or companions we examined:
1) the Hipparcos photometry,
2) the stellar chromospheric activity indicators like the EW of sensitive lines,
and 3) spectral line bisectors measurement.

\begin{table}
\begin{center}
\caption{Orbital parameters of the RV measurements for $\alpha$ Ari b.}
\begin{tabular}{cc}
\hline
\hline
    Parameter                            & Value             \\

\hline
    Period [days]                        & 380.8  $\pm$ 0.3       \\
    $\it T$$_{\mathrm{periastron}}$ [JD] & 2451213.52 $\pm$ 7.36  \\
    $\it{K}$ [m s$^{-1}$]                & 41.1  $\pm$ 0.8        \\
    $\it{e}$                             & 0.25  $\pm$ 0.03       \\
    $\omega$ [deg]                       & 89.16 $\pm$ 6.84       \\
    $f(m)$ [$\it M_{\odot}$]                 & (2.476) $\times$ 10$^{-9}$ \\
    $m$ sin $i$ [$\it M_{Jup}$]              & 1.8 $\pm$ 0.2      \\
    $\it a$ [AU]                         & 1.2                   \\
    $\sigma$ (O-C) [m s$^{-1}$]          & 17.8                  \\

\hline

\end{tabular}
\end{center}
\end{table}

\subsection{Hipparcos photometry}

 We analyzed the Hipparcos photometry for $\alpha$ Ari in order to
search for possible brightness variations due to the rotational
modulation of cool surface spots in case the RV variation is
caused by the rotation. For 2.5 years, the Hipparcos satellite obtained 66
photometric measurements of $\alpha$ Ari, and we used the whole data
except one measurement with a large error.

\begin{figure}
   \centering
   \includegraphics[width=8cm]{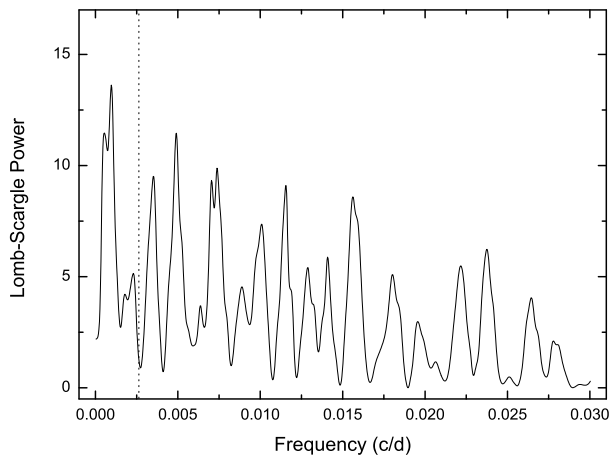}
      \caption{The Lomb-Scargle periodogram of the Hipparcos photometric measurements of
      $\alpha$ Ari for 2.5 years. The vertical dotted line marks the location of the orbital frequency.
          }
     \label{Hip_p}
   \end{figure}

Figure~\ref{Hip_p} shows the Lomb-Scargle periodogram of Hipparcos photometric
measurements calculated in the frequency interval of 0 -- 0.03 c d$^{-1}$.
There are no significant peaks near the frequency of $f_{1}$ = 0.00263 c d$^{-1}$
that corresponds to the period of 381 days.
Peaks in the periodogram are the typical low-frequency noise peaks
so they cannot be considered as significant.

\subsection{Chromospheric activity indicators}

\begin{figure}
   \centering
   \includegraphics[width=8.4cm]{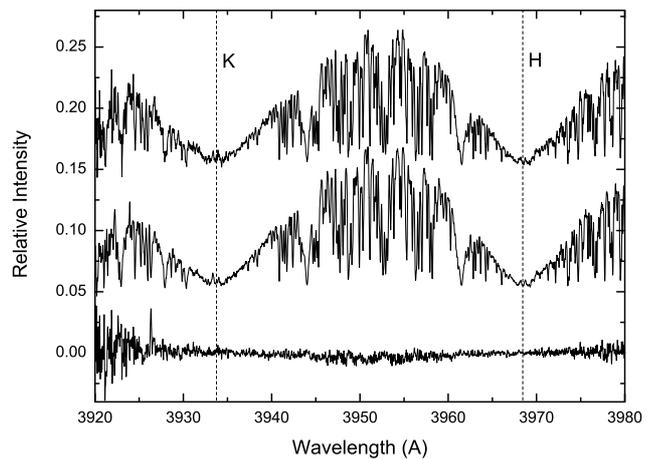}
      \caption{The spectral region of the Ca II H \& K lines in $\alpha$ Ari.
      Two spectra exhibiting the positive (top) and the negative (middle)
      RV deviations are displayed (see text). Bottom shows the differential spectrum.
      Any systematic difference in line profiles and core reversals are not apparent
      around the Ca II H \& K central region, indicating modest activity.
          }
     \label{CaII}
   \end{figure}

The EW variations of Ca II H \& K, H$_{\alpha}$, and Ca II 6882 ${\AA}$ lines
are frequently used for chromospheric activity analysis.
The emissions in the Ca II H \& K core are formed in the
chromosphere and show a typical central reversal in the existence of
chromospheric activity (Pasquini et al. 1988; Saar \& Donahue 1997).
Because the Ca II H \& K line region of the BOES spectra does not have
enough S/Ns to estimate EW variations of each spectrum reliably,
we used the averaged Ca II H \& K line profiles at the upper and lower
part of RV curve to see any systematic difference related to the RV
variations. We selected spectra for $\alpha$~Ari that correspond to
the positive deviation on JD-2453779.9451 (RV = 52.9 m s$^{-1}$) and the
negative deviation on JD-2454027.1307 (RV = -- 48.1 m s$^{-1}$) of
RV curve. These spectra in the Ca II H \& K
region were plotted in the Figure~\ref{CaII}.
Core reversal in the Ca II H \& K central region
is not obvious and there is no systematic difference in the line
profiles at these two extremal parts of the RV curve. It
means that K-giant $\alpha$~Ari, at the moment of observations, exhibits
a modest chromospheric activity.

\begin{figure}
   \centering
   \includegraphics[width=8cm]{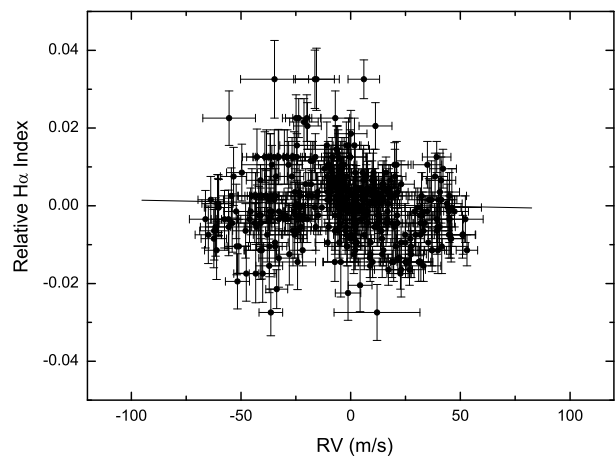}
      \caption{Relative EW(${\AA}$) vs. RV of the H$_{\alpha}$ line for $\alpha$ Ari
      from September 2004 to February 2010.
      The diagram shows no correlation between the EWs and the RVs.
          }
     \label{EW2}
   \end{figure}

The Balmer H$_{\alpha}$ line can also be a useful indicator of chromospheric activity
to determine whether the RV variation is due to intrinsic stellar
variability (K{\"u}rster et al. 2003).
Unlike in main sequence stars where H$_{\alpha}$ line is blended with telluric
and broad lines, it is easy to measure the EW of H$_{\alpha}$ line in giant stars.
The mean EW of the H$_{\alpha}$ line in $\alpha$~Ari
is measured to be 1087.5 $\pm$ 9.1 m{\AA}. The rms of 9.1 m{\AA}
corresponds to a 0.84 $\%$ variation in the EW.
Figure \ref{EW2} shows the EW of the H$_{\alpha}$ line
as a function of RV. A gradual slope (solid line) of the fitted line
shows that the EWs of the H$_{\alpha}$ line are not correlated with RV variations.

We conclude that the chromospheric activity and surface spots, if present,
have too small amplitude to be resolved in the Ca\,II H \& K  and the
H$_{\alpha}$ line EW variations.
Another activity indicator Ca II 8662 ${\AA}$ line is not
suitable because the significant fringing and saturation of our CCD spectra
at wavelengths are longer than 7500 ${\AA}$.


\subsection{Line bisector variations}

    Queloz et al. (2001) report that the analysis of the line bisector
can distinguish the origin of the observed RV variations, which is very
useful for the messy K-giant stars. The RV variations due to planetary
companions should not produce any changes in the spectral line shape,
whereas the surface inhomogeneities do. Stellar rotation modulations of
surface features can create variable asymmetries in the spectral line
profiles. Thus, the analysis of the shapes of spectral lines may prove
or disprove the existence of planetary companions.

Bisector velocity span (BVS) is defined as the difference between
two regions of the bisector at the top and bottom of the profile
to measure changes with high accuracy.
To search for variations in the spectral line shapes,
we measured the BVS using the mean profile of the spectral lines.
We calculated the mean profile of spectral lines using
the least squares deconvolution (LSD) technique (Glazunova et
al. 2008). We used the Vienna Atomic Line Database (VALD; Piskunov et
al. 1995) to prepare the list of spectral lines. The total of 4339 lines
within the wavelength region of 4500 -- 4900 and 6450 -- 6840 {\AA} were
used to construct the LSD profile, which excluded spectral
regions around the I$_{2}$ absorption region, hydrogen lines, and regions
with strong contamination by terrestrial atmospheric lines.
We estimated the BVS of the mean profile between two different flux levels
of central depth levels 0.8 and 0.25 as the span points.

\begin{figure}
   \centering
    \includegraphics[width=8cm]{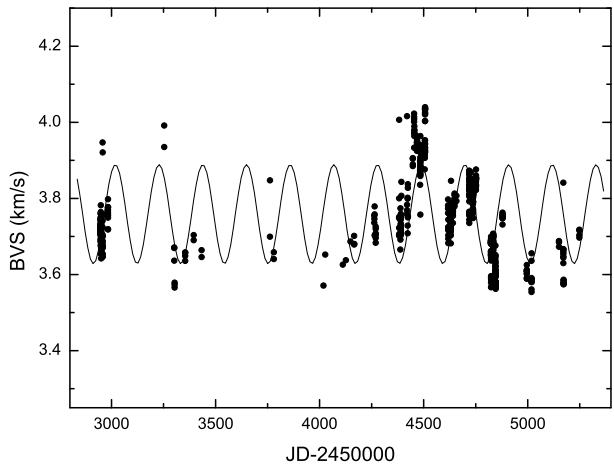}
      \caption{
      BVS variations for $\alpha$ Ari from November 2003 to February 2010.
      The solid line shows a sine wave with a period of 211 days fit.
          }
     \label{RV+BVS1}
   \end{figure}

The BVS vs JD shown in Figure~\ref{RV+BVS1} exhibits some periodic variations.
The BVSs of the JD 2454453 -- 2454506 spectra seem to be shifted with respect
to the preceding and subsequent data by the amount grater than the +100 m
s$^{-1}$. We suspect that this bias was come from the temporary mis-alignments
of the BOES echelle spectrometer. But, we did not find any BVS shift in
the K-giant $\gamma^{1}$ Leo for the same observing epoch as where a
planetary companion system was detected by using the same echelle spectrometer
BOES (Han et al. 2010). The RV code  RVI2CELL takes any changes into account
in the instrumental profile in calculations of RVs,
and thus instrumental misalignment is not likely to have affected the RV data.

\begin{figure}
   \centering
    \includegraphics[width=8cm]{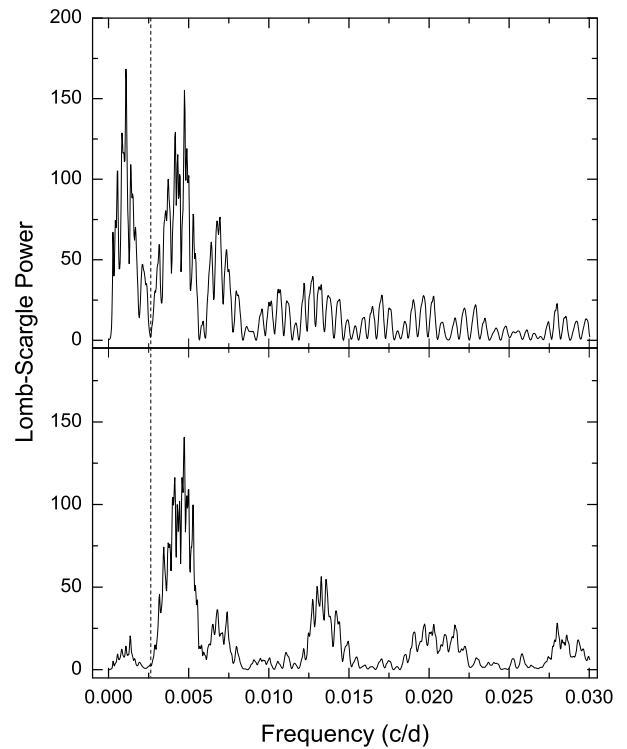}
      \caption{
      The Lomb-Scargle periodogram of the BVS variations from November 2003 to February 2010 (top panel)
      and after subtracting the low-frequency ($f_{1}$ = 0.00108 c d$^{-1}$) variations (bottom panel).
      The highest frequency is at $f_{2}$ = 0.00474 c d$^{-1}$ ($P$ = 211.2 days).
      The vertical dotted line marks the location of the orbital frequency.
          }
     \label{RV+BVS2}
   \end{figure}

To search for periodic signals in the BVS data we calculated the Lomb-Scargle
periodogram of the BVS (top panel in Figure~\ref{RV+BVS2}).
The largest peak in the Lomb-Scargle periodogram is at a very low frequency,
$f_{1}$ = 0.00108 c d$^{-1}$ ($P$ = 926.8 days) with an FAP less than 9 $\times$ $10^{-6}$.
After removal of this signal, we plot the Lomb-Scargle periodogram of
the residuals at bottom panel in Figure~\ref{RV+BVS2}.
The highest peak at $f_{2}$ = 0.00474 c d$^{-1}$ ($P$ = 211.2 days) has high FAP $\sim$ 2 $\times$ $10^{-5}$.
Figure~\ref{RV+BVS1} shows the individual fit of the frequency to BVS data.

Taking the stellar radius and rotational velocity ($v_{rot}$ sin $i$) value into account
we derived the expected range of rotational period of 172 -- 335 days.
Found about 211 day periods of BVS variations well fit this interval.
We can summarize that the variation can be attributed to
stellar surface active regions, inhomogeneities, and rotation.



\section{Discussion and conclusion}

    Our analysis of the RV measurements for $\alpha$ Ari revealed
a 381 day periodic variation that persisted between 2003 -- 2010
for almost six cycles with no  changes in phase or amplitude. We also
confirmed the short-period oscillations previously found in $\alpha$ Ari
(Kim et al. 2006). We found no evidence of the correlation between
the 381 day RV variation and the EWs of H$_{\alpha}$ lines
or the Hipparcos photometry. But, we found about 211 day periodic variations
in the BVS that we attribute to the rotational modulation of line profiles
from the surface spots and about 927 day variations in the unknown origin.
Thus, we conclude that the planetary companion in $\alpha$\,Ari is
the most likely cause of the 381 day RV variations.

If confirmed, the determined orbit of the 1.8 $M_{Jup}$ mass planet is fairly eccentric,
$e$ = 0.25 $\pm$ 0.03, but, up to now,
6 out of 22 companions around K-giants show more than 0.2 eccentricity:
    $e$ = 0.7 of $\iota$ Dra (Frink et al. 2002),
    $e$ = 0.2 of K1 III star HD 47536 (Setiawan et al. 2003),
    $e$ = 0.27 of K1 II-III star HD 13189 (Hatzes et al. 2005),
    $e$ = 0.43 of K1 III star 4 UMa (D{\"o}llinger et al. 2007),
    $e$ = 0.38 of K1.5 III star 42 Dra (D{\"o}llinger et al. 2009), and
    $e$ = 0.46 of K2 III star HD 110014 (de Medeiros et al. 2009).
The eccentricity for the planetary orbit of $\alpha$ Ari is therefore not
unusual among K-giants, and such eccentric modulation is difficult to produce
by surface variations or stellar oscillations.

The source $\alpha$ Ari belongs to yet smaller group of planet hosting
K-giants showing short-period oscillations (e.g. $\beta$\,Gem (Hatzes \&
Zechmeister 2007), $\iota$ Dra (Zechmeister et al. 2008)).


\begin{acknowledgements}
      Support for MGP was provided by the National Research Foundation of Korea
      to the Center for Galaxy Evolution Research.
      We thank the developers of the Bohyunsan Observatory Echelle Spectrograph (BOES) and
      all staff of the Bohyunsan  Optical Astronomy Observatory (BOAO).

\end{acknowledgements}
%


\end{document}